\documentclass[aps,prl,
reprint,groupedaddress,nofootinbib,longbibliography,floatfix]{revtex4-1}
\usepackage[T1]{fontenc}
\usepackage{lmodern}
\usepackage{amsmath,amssymb,bm}
\usepackage{graphicx}
\usepackage{xcolor}
\definecolor{paperblue}{RGB}{0,72,153}
\usepackage[colorlinks=true,linkcolor=paperblue,citecolor=paperblue,urlcolor=paperblue]{hyperref}
\newcommand{\dd}{\mathrm d}
\newcommand{\dr}{\mathrm{dr}}
\newcommand{\KL}{D_{\mathrm{KL}}}
\newcommand{\F}{\mathcal F}
\newcommand{\E}{\mathcal E}
\newcommand{\Sec}[1]{\emph{\color{paperblue}#1.---}}

\begin{document}

\title{Rare-History Transitions in Temporally Random Integrable Quantum Circuits}
\author{Tingfei Li\textsuperscript{1,3,4}}
\email{tfli@hbu.edu.cn}
\author{Jie Gu\textsuperscript{2}}
\email{jiegu1989@gmail.com}
\affiliation{\textsuperscript{1}College of Physics Science and Technology, Hebei University, Baoding 071002, China\\
\textsuperscript{2}Chengdu Academy of Educational Sciences, Chengdu 610036, China\\
\textsuperscript{3}Hebei Key Laboratory of High-precision Computation and Application of Quantum Field Theory, Baoding 071002, China\\
\textsuperscript{4}Hebei Research Center of the Basic Discipline for Computational Physics, Baoding 071002, China}
\date{\today}
\begin{abstract}
We study current fluctuations in a temporally random integrable quantum circuit. Commutativity reduces every drive history exactly to its layer composition, turning annealed fluctuations into a competition between current gain and the large-deviation cost of rare compositions. For each fixed composition, homogeneous thermodynamic Bethe ansatz dressing supplemented by ballistic fluctuation theory yields the conditional current statistics. Their annealed large-deviation contraction predicts a first-order switch between two dominant history classes, terminating on a line of regular cusp endpoints. Finite-time analysis shows how the switch is rounded. Thus temporal randomness can act as an emergent order-parameter-like coordinate in trajectory space.
\end{abstract}

\maketitle

\Sec{Introduction}
Transport is often summarized by an average current, but rare fluctuations can reveal structure invisible at the level of mean behavior. Full counting statistics (FCS) records atypical transferred charge or spin, while large-deviation theory organizes its long-time generating function as a trajectory-space free energy~\cite{Levitov1993,Klich2003,Schonhammer2007,KlichLevitov2009,Esposito2009,Touchette2009,JackSollich2010,PerfettoGambassi2020,BertiniFCS2023,McCulloch2023}. Nonanalyticities of this object can signal dynamical phase transitions between distinct classes of trajectories even when the unbiased dynamics is smooth~\cite{GarrahanLesanovsky2010,Lesanovsky2013,Macieszczak2016}. Random quantum circuits provide a controlled setting for such questions~\cite{Nahum2017,Nahum2018,Keyserlingk2018,Rakovszky2018,Chan2018,ZhouNahum2019,Fisher2023}.

Integrability offers complementary control through commuting transfer matrices, stable quasiparticles, and thermodynamic Bethe ansatz~\cite{Yang1967,Baxter1972,BaxterBook,KorepinBook,Faddeev1996,YangYang1969,TakahashiSuzuki1972,TakahashiBook}. Genuine temporal randomness usually conflicts with this structure because generic time-dependent interacting layers do not commute. Integrable Floquet and Trotter circuits avoid this obstruction in deterministic settings~\cite{GritsevPolkovnikov2017,Vanicat2018,LjubotinaCircuit2019,Aleiner2021,Vernier2023,MiaoVernier2023,Miao2024,Paletta2025,Hubner2025}, while Ref.~\cite{Wang2026} recently constructed a genuinely time-random circuit in which every temporal sequence preserves the same Yang--Baxter structure. This makes the model unusually well suited to rare-history questions: temporal randomness is retained without sacrificing access to the thermodynamic quasiparticle description of interacting integrable transport. That work focused on typical-sequence transport. Here we ask the complementary question: how do {rare temporal histories themselves} enter the current FCS?

An atypical current can arise either from an unusual many-body fluctuation at a typical drive composition or from an atypical temporal composition that changes the quasiparticle dynamics and lowers the current cost. The circuit of Ref.~\cite{Wang2026} makes this competition tractable: because its two random layers belong to one commuting transfer-matrix family, every length-$t$ temporal word reduces exactly to its empirical layer fraction $\eta=N_1/t$, where $N_1$ is the number of type-1 layers. For Bernoulli sampling with mean $\bar\eta$, the annealed problem therefore balances the fixed-composition scaled cumulant generating function (SCGF) $\mu(s,\eta)$, with counting field $s$, against the Bernoulli relative-entropy cost $\KL(\eta\Vert\bar\eta)$. For fixed $\eta$ the circuit is spatially homogeneous: thermodynamic Bethe ansatz (TBA) supplies the occupations, scattering, and dressing structure, while ballistic fluctuation theory (BFT) generates the finite-$s$ deformation and hence $\mu(s,\eta)$~\cite{DoyonMyers2020,Myers2020,PerfettoDoyon2021,DoyonBMFT2023}. The annealed SCGF is then obtained by optimizing over temporal compositions.

At the minimal interacting root $P=3$ (here $P$ labels the root-of-unity truncation used below), the annealed objective develops two macroscopically separated maxima that become equal and exchange global dominance as the counting field is varied. The selected composition and biased current jump simultaneously, yielding a first-order rare-history saddle transition. Varying $(\bar\eta,s,\Delta u)$, where $\Delta u$ is the spectral-parameter control defined below, extends the switch to a coexistence surface whose regular endpoints are described by a cusp normal form, with regularized scalings consistent with the expected $1/2$ and $3/2$ powers. At finite observation time the singular switch is rounded into a two-saddle crossover with width $\delta s\sim[t|\Delta j|]^{-1}$, where $\Delta j$ denotes the jump in the biased current across coexistence.

These results show that temporal randomness can act not merely as external noise but as an order-parameter-like coordinate that reorganizes current large deviations. The exact history reduction, the conditional homogeneous TBA/BFT calculation, and the probabilistic annealed contraction are logically distinct. Accordingly, the transition established here is a saddle transition of the annealed large-deviation objective rather than a finite-size spectral theorem. Its order-parameter-like variable is the temporal composition $\eta$, not a spatial hydrodynamic field.


\Sec{Circuit and exact history reduction}
We consider a spin-$\tfrac12$ chain of length $L$, with conserved local spin
$q_j=\sigma_j^z/2$.
Let $\mathcal T(z)$ be the row-to-row transfer matrix of the period-three inhomogeneous six-vertex model used in Ref.~\cite{Wang2026}, where $z$ is the complex spectral parameter. In the easy-plane regime the unitary line is imaginary, so we use real rapidity coordinates $u$ and define $T(u):=\mathcal T(iu)$. The Yang--Baxter equation gives $[T(u),T(v)]=0$ for all real $u,v$. Two unitary circuit layers are obtained from ratios of this family,
\begin{equation}
 U_0=T^{-1}(u_1)T(u_2),\qquad
 U_1=T^{-1}(u_3)T(u_1),
\label{eq:layers}
\end{equation}
where the transfer matrices at the displayed arguments are assumed invertible. It follows algebraically that $[U_0,U_1]=0$. For a binary temporal word $\omega=(w_1,\ldots,w_t)$, define $N_1(\omega)=\sum_k w_k$. Reordering the commuting factors gives
\begin{align}
 U_\omega&=U_{w_t}\cdots U_{w_1}
 =U_0^{t-N_1}U_1^{N_1} \nonumber\\
 &=T(u_2)^{t-N_1}T(u_1)^{-t+2N_1}T(u_3)^{-N_1}.
\label{eq:word}
\end{align}
Eq.~\eqref{eq:word} is an exact finite-time, finite-size operator identity. 
The ordering of the word disappears and only its composition $\eta=N_1/t$ remains.
The microscopic model is therefore the circuit of Ref.~\cite{Wang2026}, restricted below to an easy-plane root of unity and a spectral-parameter slice.
On this slice we keep $u_1-u_2$ fixed and use
$\Delta u=u_3-u_1$ as a control parameter.

The current observable is defined after the thermodynamic limit on an infinite chain. Let $A = (-\infty, 0]$ with associated spin $Q_A$. Define $J_t = Q_A(0) - Q_A(t)$ as the spin transferred across the interface at the origin in a two-time measurement protocol. We use the zero-field infinite-temperature tracial state, represented at finite size by $\varrho_0=\mathbb I/2^L$ before taking $L\to\infty$. In particular $[\varrho_0,Q_A]=0$. The moment-generating function for a word is
\begin{equation}
 Z_\omega(s,t)=\operatorname{Tr}\!\left[e^{-sQ_A}U_\omega e^{sQ_A}
 \varrho_0U_\omega^\dagger\right].
\label{eq:mgf}
\end{equation}
It depends on $\omega$ only through $N_1$. We denote the common value for all length-$t$ words with the same $N_1$ by $Z_{t,N_1}(s)$. We then define the fixed-composition SCGF by
\begin{equation}
 \mu(s,\eta)=\lim_{\substack{t\to\infty\\N_1/t\to\eta}}
 \frac1t\log Z_{t,N_1}(s),\qquad \mu(0,\eta)=0,
\label{eq:conditional}
\end{equation}
 Its derivatives generate the conditional scaled cumulants. In particular $j(s,\eta)=\partial_s\mu(s,\eta)$ is the current in the $s$-biased state.

For independent Bernoulli layers with $\Pr(w_k=1)=\bar\eta$, the annealed generating function can be grouped exactly by the composition $N_1/t$:
\(
\mathbb E_\omega Z_\omega(s,t)
=\sum_{N=0}^t\binom tN
\bar\eta^N(1-\bar\eta)^{t-N}Z_{t,N}(s).
\)
For $0<\bar\eta<1$, the Bernoulli relative entropy is
\(
\KL(\eta\Vert\bar\eta)
=
\eta\log\frac{\eta}{\bar\eta}
+(1-\eta)\log\frac{1-\eta}{1-\bar\eta}.
\)
The binomial weight obeys Cram\'er's theorem,
$\Pr(N_1/t\simeq\eta)\asymp e^{-t\KL(\eta\Vert\bar\eta)}$,
while Eq.~\eqref{eq:conditional} gives
$Z_{t,N}(s)\asymp e^{t\mu(s,\eta)}$ at fixed $\eta=N/t$.
Assuming the conditional limit is sufficiently uniform in $\eta$, the Laplace principle therefore yields
\begin{equation}
\mu_{\rm ann}(s;\bar\eta)
\equiv
\lim_{t\to\infty}\frac1t
\log\mathbb E_\omega Z_\omega(s,t)
=
\sup_{0\leq\eta\leq1}\F(s,\eta;\bar\eta),
\label{eq:contraction}
\end{equation}
where
$\F(s,\eta;\bar\eta)=\mu(s,\eta)-\KL(\eta\Vert\bar\eta)$.
Unless stated otherwise, the dependence of $\mu$ and $\F$ on the fixed spectral parameters, including $\Delta u$, is suppressed. Thus annealing optimizes the current-generating gain over atypical drive compositions, penalized by their large-deviation cost. If $I_\eta(j)$ denotes the fixed-composition current rate function and $I_{\rm ann}(j)$ its annealed counterpart, then equivalently
\(
I_{\rm ann}(j)
=
\inf_\eta\!\left[
\KL(\eta\Vert\bar\eta)+I_\eta(j)
\right].
\)

\begin{figure*}[htb!]
\includegraphics[width=0.7\textwidth]{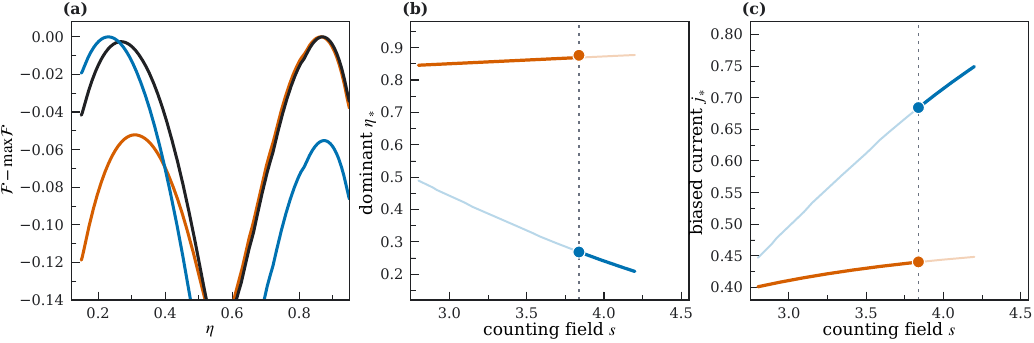}
\caption{{Ballistic-fluctuation evidence for a first-order rare-history switch at $P=3$.}
(a) Objective near coexistence for $\bar\eta=0.8$, $u_1-u_2=4.1$, and $\Delta u=u_3-u_1=1$. The vertical axis is shifted by the instantaneous maximum,
$\F(s,\eta;\bar\eta)-\max_{\eta}\F(s,\eta;\bar\eta)$.
Orange, black, and blue denote fields below, at, and above the coexistence field $s_c$.
(b) Shape-preserving interpolation of the locally refined peak positions. The saturated segments show the globally selected branch, while the faint continuations are metastable local-maxima branches. The vertical dotted line marks $s=s_c$.
(c) For the globally selected composition $\eta_*(s)$,
$j_*(s)\equiv j(s,\eta_*(s))=\partial_s\mu(s,\eta_*(s))$. The dotted line again marks $s=s_c$. At coexistence both competing branches are shown.
A Gaussian composition regulator of width $\sigma_\eta=0.004$ suppresses grid-scale teeth. The switch also appears at $\sigma_\eta=0$; see Supplemental Material~\cite{SM}.}
\label{fig:firstorder}
\end{figure*}

\Sec{TBA and ballistic-fluctuation evaluation of $\mu(s,\eta)$}
Fixing the temporal composition $\eta$ removes the randomness of the drive: the remaining problem is a spatially homogeneous, translation-invariant integrable circuit. There is no slowly varying spatial profile or $x,t$-dependent root density to evolve. We describe the homogeneous state directly as a gas of interacting Bethe quasiparticles, labeled by a species index $a$ and rapidity $\lambda$. The thermodynamic Bethe ansatz (TBA) specifies their occupations, scattering, and dressing structure; the fixed-$\eta$ quasienergy supplies their propagation data; and ballistic fluctuation theory (BFT) describes how the occupations change when the current is biased by the counting field $s$~\cite{TakahashiSuzuki1972,TakahashiBook,DoyonMyers2020,Myers2020,PerfettoDoyon2021,DoyonBMFT2023}. At the root-of-unity points considered here there are finitely many quasiparticle species. We denote the Takahashi--Suzuki species data for branch $a$ by $(n_a,v_a)$ and the associated model-specific kernel by $\mathcal K_n^v(\lambda)$. Their explicit forms are given in Ref.~\cite{SM}.

Interactions renormalize the charge and propagation of each quasiparticle through the standard TBA dressing operation. For any bare one-particle quantity $h_a(\lambda)$,
\begin{equation}
 h_a^{\dr}
 =
 h_a-\sum_b\sigma_b A_{ab}\star
 \bigl(\vartheta_b h_b^{\dr}\bigr),
\label{eq:dressing}
\end{equation}
where $\vartheta_a(\lambda)$ is the occupation of mode $(a,\lambda)$, $A_{ab}$ is the two-body scattering kernel, $\sigma_b$ is the root-of-unity sign convention, and $(f\star g)(\lambda)=\int\dd\alpha\,f(\lambda-\alpha)g(\alpha)$. Thus ``dressing'' is simply the many-body renormalization that converts bare quasiparticle data into the effective quantities carried in a finite-density state.

The model-specific input is how the history composition changes the quasiparticle dispersion. Because the two circuit layers commute, their quasienergies add, so the bare quasienergy derivative
$\epsilon'_{a,\eta}(\lambda)\equiv\partial_\lambda\epsilon_{a,\eta}(\lambda)$
is affine in $\eta$ and directly defines the homogeneous bare propagation data:
\begin{equation}
\begin{aligned}
 \epsilon'_{a,\eta}(\lambda)=2\pi\bigl\{&
 (1-2\eta)\mathcal K_{n_a}^{v_a}(\lambda-u_1)
 -(1-\eta)\mathcal K_{n_a}^{v_a}(\lambda-u_2)\\
 &+\eta\mathcal K_{n_a}^{v_a}(\lambda-u_3)
 \bigr\}.
\end{aligned}
\label{eq:source}
\end{equation}
This is the microscopic bridge from a temporal history to transport: changing $\eta$ changes the quasiparticle propagation entering the current. For rational $\eta=r/q$, Eq.~\eqref{eq:source} is the quasienergy per step of the commuting supercell $U_0^{q-r}U_1^r$. Irrational $\eta$ is defined by rational approximation. We take the infinite-chain single-interface limit before $t\to\infty$ with $N_1/t\to\eta$, assuming the regularity in $\eta$ required by Eq.~\eqref{eq:contraction}.

The counting field $s$ biases the ensemble toward atypical currents. In the quasiparticle description this continuously deforms the occupations $\vartheta_a(\lambda;s)$. Writing
$\vartheta_a=(1+e^{\E_a})^{-1}$, which defines $\E_a(\lambda;s)$, and taking the bare magnetization to be $m_a=n_a$, the BFT flow and the corresponding biased current are
\begin{equation}
\begin{aligned}
 \partial_s\E_a(\lambda;s)
 &=
 -\sigma_a\operatorname{sgn}\!\left[
 (\epsilon'_{a,\eta})^{\dr}(\lambda;s)
 \right]m_a^{\dr}(\lambda;s),\\
 j(s,\eta)
 &=
 \sum_a\int\frac{\dd\lambda}{2\pi}\,
 \sigma_a\epsilon'_{a,\eta}(\lambda)
 \vartheta_a(\lambda;s)m_a^{\dr}(\lambda;s).
\end{aligned}
\label{eq:bft}
\end{equation}
The sign selects the physical propagation direction, while $m_a^{\dr}$ is the magnetization carried by the dressed quasiparticle. Starting from the infinite-temperature state at $s=0$, we evolve Eq.~\eqref{eq:bft}, recompute the dressed quantities in the instantaneous biased state, and obtain the conditional SCGF from
$\partial_s\mu(s,\eta)=j(s,\eta)$ with $\mu(0,\eta)=0$. In short,
$\eta$ fixes the homogeneous quasiparticle data, TBA dressing accounts for their many-body renormalization, BFT gives the biased current, and integrating that current gives $\mu(s,\eta)$. The composition dependence in Eq.~\eqref{eq:source} is exact, whereas the finite-$s$ evaluation through Eq.~\eqref{eq:bft} is the thermodynamic ballistic-fluctuation approximation used below. Here ``Euler scale'' refers only to this asymptotic ballistic description of current statistics, not to a spatially inhomogeneous hydrodynamic evolution.

A simple $P=3$ result shows analytically why this procedure can generate the saddle competition discussed later. Spin-reversal symmetry gives $\mu(s,\eta)=\mu(-s,\eta)$. With $u_2=0$, $u_1=\ell$, $u_3=\ell+\Delta u$, define
$b_2(x)=\sqrt3/[2\pi(\cosh2x+1/2)]$ and
$h_\eta(\lambda)=-(1-\eta)b_2(\lambda)+(1-2\eta)b_2(\lambda-\ell)+\eta b_2(\lambda-\ell-\Delta u)$.
Zero-bias BFT then gives
\begin{equation}
\begin{aligned}
 \mu(s,\eta)
 &=\frac12 c_2(\eta,\Delta u)s^2+O(s^4),\\
 c_2(\eta,\Delta u)
 &=\frac34\int_{-\infty}^{\infty}\dd\lambda\,|h_\eta(\lambda)|.
\end{aligned}
\label{eq:c2}
\end{equation}
Although $h_\eta$ is affine in $\eta$, the absolute value makes $c_2(\eta,\Delta u)$ nonlinear in $\eta$ at fixed $\Delta u$. This provides an explicit mechanism by which the current-generating gain can compete with the convex Bernoulli cost in Eq.~\eqref{eq:contraction}. The finite-$s$ phase diagram nevertheless uses the full BFT flow in Eq.~\eqref{eq:bft}, not this quadratic expansion.

Numerically, rapidity is discretized with $N_\lambda$ points on the periodic box $[-14,14)$, dressing convolutions are evaluated by FFT, and the BFT flow is integrated with a second-order Heun method. Convergence and the remaining numerical controls are documented in Ref.~\cite{SM}.

\Sec{History coexistence}
An interior locally stable history satisfies $\partial_\eta\F=0$ and $\partial_\eta^2\F<0$. We use
$\operatorname{logit}x\equiv\log[x/(1-x)]$.
Since
$\partial_\eta\KL=\operatorname{logit}\eta-\operatorname{logit}\bar\eta$,
the saddle equation can be written as
\begin{equation}
 \operatorname{logit}\eta_*=\operatorname{logit}\bar\eta
 +\partial_\eta\mu(s,\eta_*).
\label{eq:saddle}
\end{equation}
We denote by $\eta_-$ and $\eta_+$ the two locally stable histories that compete for the global supremum in \eqref{eq:contraction}, ordered so that $\eta_-<\eta_+$.
Two stable compositions $\eta_-<\eta_+$ coexist at the counting field $s_c$ only when
\begin{equation}
 \F(s_c,\eta_-;\bar\eta)=\F(s_c,\eta_+;\bar\eta).
\label{eq:coexistence}
\end{equation}
We define the corresponding branch currents by
$j_\pm\equiv j(s_c,\eta_\pm)$
and their jump by
$\Delta j\equiv j_+-j_-$.
This equal-height condition is essential: multiple stationary points, velocity rearrangements, or root-density features alone do not establish a phase transition.

Numerically, at each $s$ we evaluate $\F$ on a uniform composition grid of $N_\eta$ points, smooth only along $\eta$ with a Gaussian of fixed physical width $\sigma_\eta$ to suppress grid-scale teeth, and refine each macroscopic maximum by a local spline. Coexistence is located from a sign change of the height difference between the two tracked branches. Candidates are rejected if the branch separation falls below $0.08$, branch identities are discontinuous, the equal-height residual is too large, or a third maximum is globally higher.

At the minimal interacting root $P=3$, Fig.~\ref{fig:firstorder}(a)--(c) shows the representative $\bar\eta=0.8,\Delta u=1$ section. The two maxima are equal at $s_c=3.83821$, with $(\eta_-,\eta_+)=(0.26845,0.87649)$. The selected composition changes by $0.60804$, while the biased current changes by $|\Delta j|=0.24408$. Both discontinuities are macroscopic relative to the grid and smoothing scales. The current bias is therefore realized by switching between two exponentially distinct classes of random temporal words. Within Euler-scale BFT, this is a nonanalytic saddle switch of the limiting annealed SCGF. Independent scans show the same switch already at $\sigma_\eta=0$ and convergence with $N_\lambda$ and $N_\eta$~\cite{SM}.

Differentiating Eq.~\eqref{eq:coexistence} gives a trajectory-space Clapeyron relation. At fixed $\Delta u$, stationarity removes the implicit derivatives of $\eta_\pm$, while $\partial_s\F=j$ and $\partial_{\bar\eta}\F=(\eta-\bar\eta)/[\bar\eta(1-\bar\eta)]$. Hence
\begin{equation}
 \frac{\dd s_c}{\dd\bar\eta}=
 \frac{\eta_+-\eta_-}{\bar\eta(1-\bar\eta)[j(s_c,\eta_-)-j(s_c,\eta_+)]}.
\label{eq:clapeyron}
\end{equation}
Across the computed $P=3$ coexistence sections, this identity agrees with finite differences of $s_c(\bar\eta)$ to median relative errors between $1.1\times10^{-4}$ and $3.8\times10^{-4}$. It provides a nonlocal consistency check linking the phase-boundary slope to two independently extracted jumps.

\begin{figure}[tb!]
\includegraphics[width=\columnwidth]{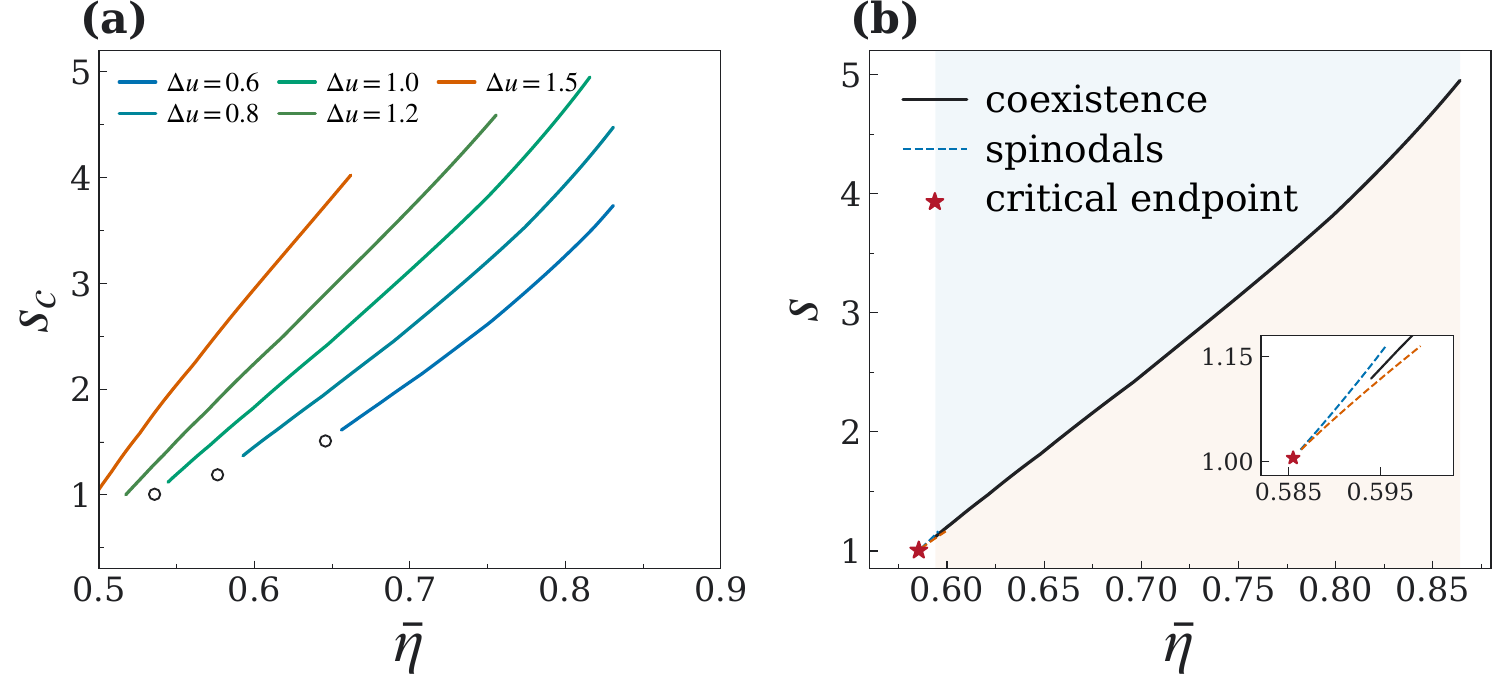}
\caption{{Regularized $P=3$ TBA/BFT trajectory phase diagram.}
(a) Projection of equal-height BFT sections onto the $(\bar\eta,s_c)$ plane, with color identifying $\Delta u$. Open circles mark the endpoints of the fixed-$\Delta u$ coexistence branches obtained by continuation.
(b) The $\Delta u=1$ slice: the solid black curve is coexistence, dashed curves delimit the spinodal wedge, and the star marks the critical endpoint extracted from the regularized data. The inset enlarges the neighborhood of this endpoint. Pale blue and orange identify the globally selected low- and high-$\eta$ history sectors, respectively, not equilibrium phases.}
\label{fig:phase}
\end{figure}

\Sec{Coexistence surface and criticality}
Varying $(\bar\eta,s,\Delta u)$ extends the switch to a two-dimensional coexistence surface (Fig.~\ref{fig:phase}). A spinodal obeys $\partial_\eta \F=\partial_{\eta}^2\F=0$. At a regular critical endpoint the two maxima and the intervening minimum merge, so additionally $\partial_\eta^3 \F=0$ and $\partial_\eta^4 \F<0$. Denote the control vector by
$\bm g=(\bar\eta,s,\Delta u)$
and its endpoint value by
$\bm g_{\rm E}=(\bar\eta_{\rm E},s_{\rm E},\Delta u_{\rm E})$.
Let $\eta_{\rm E}$ be the coalesced composition and define
$\F_{\rm E}\equiv
\F(s_{\rm E},\eta_{\rm E};\bar\eta_{\rm E})
\big|_{\Delta u=\Delta u_{\rm E}}$.
For
$\delta\bm g\equiv\bm g-\bm g_{\rm E}$,
analytic local changes of the control coordinates reduce the objective to the cusp normal form
\begin{equation}
\begin{aligned}
\F-\F_{\rm E}
&=Hx+\frac{R}{2}x^2-\frac{U}{4}x^4
+O\!\left(x^5,x^2\|\delta\bm g\|^2\right),\\
x&=\eta-\eta_{\rm E}.
\end{aligned}
\label{eq:cusp}
\end{equation}
Here $H$ and $R$ are analytic combinations of $\delta\bm g$, chosen as the field-like and curvature-like cusp coordinates and vanishing at the endpoint, while $U=-\partial_\eta^4\F|_{\rm E}/6>0$. The remaining regular control direction parametrizes motion along the endpoint line and changes the normal-form coefficients only analytically. 
Stationarity gives $Ux^3-Rx-H=0$. On coexistence $H=0$, so $x_\pm=\pm\sqrt{R/U}$, while the spinodals additionally satisfy $3Ux^2-R=0$.
 Along a chosen transverse one-parameter path through the endpoint, let $\delta r$ denote its signed displacement from the endpoint. The coexistence jump $\Delta\eta\equiv\eta_+-\eta_-$ and the spinodal width $\Delta H_{\rm sp}\equiv|H_{\rm sp}^{(+)}-H_{\rm sp}^{(-)}|$ then scale as
\begin{align}
 \Delta\eta&=2\sqrt{R/U}\propto|\delta r|^{1/2},\nonumber\\
 \Delta H_{\rm sp}&=\frac{4R^{3/2}}{3\sqrt{3U}}
 \propto|\delta r|^{3/2}.
\label{eq:exponents}
\end{align}
The regularized finite-window fits reported in Ref.~\cite{SM},
$\Delta\eta\propto|\delta r|^\beta$ and
$\Delta H_{\rm sp}\propto|\delta r|^\gamma$, give
$\beta=0.515$ and $\gamma=1.514$, close to the cusp predictions
$1/2$ and $3/2$. These values are consistent with the regular cusp normal form but are not regulator-independent exponent measurements. 
With $N_\lambda=1024$ rapidity grid points and Gaussian composition
smoothing width $\sigma_\eta=0.003$, the regularized slice estimate is $(\bar\eta_{\rm E},s_{\rm E},\eta_{\rm E})=(0.5855,1.0052,0.5476)$. Geometrically, the two-dimensional coexistence surface in $(\bar\eta,s,\Delta u)$ terminates on a one-dimensional endpoint line. Fig.~\ref{fig:phase}(b) displays only its fixed-$\Delta u=1$ slice.

\Sec{Finite-time rounding of the trajectory transition}
The transition is sharp only after the long-time limit. Near a coexistence point, suppose the two maxima are nondegenerate and write their objective values as $\F_\pm(s)$. Laplace expansion of the exact binomial sum then has the two-saddle form
\begin{equation}
 \mathbb E_\omega Z_\omega(s,t)\simeq
 B_-(t,s)e^{t\F_-(s)}+B_+(t,s)e^{t\F_+(s)},
\label{eq:rounding}
\end{equation}
where $\log B_\pm=o(t)$ contains the conditional prefactor and the local saddle curvature. If the conditional prefactor is $O(1)$, the $t^{-1/2}$ from the binomial probability, the factor $t$ from replacing the sum by an integral, and the $t^{-1/2}$ from the Gaussian saddle integral cancel, so $B_\pm=O(1)$. No universal power of $t$ is assumed. At finite $t$ this expression is analytic. Since $\partial_s(\F_+-\F_-)=j_+-j_-$ at coexistence, the crossover width scales as $\delta s\sim[t|\Delta j|]^{-1}$, and the relative weights of the two history sectors have a logistic crossover. Thus the measured finite-time distribution should become bimodal in $\eta$ before the limiting SCGF develops a cusp. Moreover, the jump of $\partial_s\mu_{\rm ann}$ from $j_+$ to $j_-$ produces a Maxwell-type affine segment between those currents in the Legendre--Fenchel rate function. These consequences distinguish controlled finite-time rounding of a saddle transition from a microscopic level crossing.

\Sec{Discussion}
The essential point is that temporal disorder becomes an active thermodynamic coordinate of the biased ensemble. Yang--Baxter commutativity makes the empirical composition $\eta$ a sufficient statistic for an entire temporal word, while the current bias can favor compositions that are exponentially rare in the original drive. The competing saddles therefore represent distinct ensembles of temporal histories, not spatially coexisting states. More broadly, this suggests that annealed fluctuations in commuting random drives can reorganize the disorder ensemble itself, even when every fixed-composition circuit remains spatially homogeneous.

The resulting singularity is an asymptotic large-deviation prediction whose conditional input is evaluated at ballistic scale. Its microscopic realization should be tested using a size- and time-converged single-interface FCS calculation. Tensor-network evolution~\cite{Valli2025,Samajdar2024,Yadalam2026} and importance sampling targeted at rare temporal compositions are natural routes. Exact XXZ results for form factors, finite-temperature correlations, generalized Gibbs ensembles, and static FCS provide complementary microscopic benchmarks~\cite{Kitanine1999,Gohmann2004,Pozsgay2013,Collura2017,Calabrese2020FCS,EislerRacz2013,Belliard2018,Belliard2025}, while related studies of random integrable chains, driven systems, tilted non-Hermitian generators, and interacting automata suggest further comparison points~\cite{Essler2018,Agrawal2019,MacDonald2021,Zadnik2024,Krajnik2022,KrajnikCrit2024,Krajnik2025,GopalakrishnanFCS2024,Yoshimura2025,Yoshimura2026}. Such calculations could establish how the two history sectors emerge at finite time and determine how diffusive, quasiparticle, or superdiffusive corrections modify the approach to the ballistic limit~\cite{Sirker2009,Sirker2011,Piroli2017,DeNardis2018,DeNardis2019Diff,Gopalakrishnan2018,Ljubotina2017,Ljubotina2019,GopalakrishnanVasseur2019,DeNardis2019Anom,Ilievski2021,Bulchandani2021}.

Several extensions are immediate. Tracking the moving zeros of $(\epsilon'_{a,\eta})^{\dr}$ would remove the composition regulator and permit a regulator-free determination of the endpoint line. Extending the analysis to higher root-of-unity truncations would test whether the two-history structure persists or gives way to multiple competing sectors. Correlated temporal disorder and drives with more than two commuting layers would replace the scalar $\eta$ by a higher-dimensional composition variable, opening the possibility of richer coexistence geometries. Comparing annealed and quenched current statistics would further clarify when rare transport is controlled by reorganizing the drive rather than by fluctuations within a typical history. The broader question is thus how integrability converts temporal randomness into structure in trajectory space.

\Sec{Acknowledgments}
Tingfei Li acknowledges support from the National Natural Science Foundation of China (Grant No.~12605129). The authors thank Miao He for valuable discussions.

\bibliography{references}

\end{document}